\documentstyle[prd,aps,epsfig]{revtex}  
\begin{document}                
\title{Probing the gravitational well: No supernova explosion 
in spherical symmetry with general relativistic Boltzmann
neutrino transport}

\author{
Matthias Liebend\"{o}rfer$^{1,2,3}$, Anthony Mezzacappa$^{2}$,
Friedrich-Karl Thielemann$^{1,2}$, \\
O. E. Bronson Messer$^{2,3,4}$, W. Raphael Hix$^{2,3,4}$,
Stephen W. Bruenn$^{5}$
}

\address{
$^{1}$ Department of Physics and Astronomy, University of Basel,\\
Klingelbergstrasse 82, 4056 Basel Switzerland \\
$^{2}$ Physics Division, Oak Ridge National Laboratory, Oak Ridge, Tennessee
37831-6354 \\
$^{3}$ Department of Physics and Astronomy, University of Tennessee, Knoxville,
Tennessee 37996-1200 \\
$^{4}$ Joint Institute for Heavy Ion Research, Oak Ridge National 
Laboratory,\\
Oak Ridge, Tennessee 37831-6374\\
$^{5}$ Department of Physics, Florida Atlantic University, Boca Raton,
Florida 33431-0991
}
\maketitle


\begin{abstract}                
We report on the stellar core collapse,
bounce, and postbounce evolution of a $13$ M$_{\odot}$ star
in a self-consistent general relativistic spherically symmetric
simulation based on Boltzmann neutrino transport. We conclude
that
approximations to exact neutrino transport and the omission of general
relativistic effects were not alone responsible for the
failure of numerous preceding attempts to model supernova explosions
in spherical symmetry. Compared to simulations in Newtonian
gravity, the general relativistic simulation results in a smaller
shock radius. We however argue that the higher
neutrino luminosities and rms energies in the general relativistic
case could lead to a larger supernova explosion energy.
\end{abstract}


\section{\bf Back to the Past?}

In the pioneering days of supernova simulations (Colgate and White 
\cite{Colgate_White_66}, May and White \cite{May_White_67}) the 
consideration of general relativity (GR) was standard. The 
observable event of a stellar core collapse followed by a supernova
explosion was an ideal application for the newly derived 
Einstein equations in spherical symmetry (Misner and Sharp 
\cite{Misner_Sharp_64}). Lindquist \cite{Lindquist_66}
formulated the GR Boltzmann equation,
and Wilson \cite{Wilson_71} carried out simulations based on
GR Boltzmann neutrino transport using parametrized neutrino interactions.
This early epoch laid the foundation leading to our current
 understanding of the supernova mechanism: A collapsing stellar iron core 
bounces at nuclear densities and launches a shock wave outwards through 
the infalling outer layers. The shock is energized by neutrinos diffusing out 
of the hot proto-neutron star, which deposit a fraction of their energy 
in the shock-dissociated matter via absorption on the free nucleons.

The early models underwent refinements in many respects.
The equation of state evolved from simple polytropes to significantly more
realistic models (Baron et al.\cite{Baron_Cooperstein_Kahana_85a},
Lattimer and Swesty \cite{Lattimer_Swesty_91}).
The neutrino opacities were improved
(Tubbs and Schramm \cite{Tubbs_Schramm_75},
Schinder and Shapiro \cite{Schinder_Shapiro_82},
Bruenn \cite{Bruenn_85}), and sophisticated
multigroup flux-limited diffusion neutrino transport schemes were
developed based on this ``standard'' nuclear physics input
(Arnett \cite{Arnett_77}, Bowers and Wilson \cite{Bowers_Wilson_82},
Bruenn \cite{Bruenn_85}, Myra et al.\cite{Myra_et_al_87}).

However, improved approximations in the implementation of neutrino 
physics seemed to decrease the likelihood of successful explosions in
spherical symmetry. This was especially true with the inclusion
of the detailed neutrino energy spectrum via multigroup simulations
and for simulations that took the inelastic neutrino--electron scattering
into account (Bruenn \cite{Bruenn_89a,Bruenn_89b}). The idea
that the amount of dissociation energy expended by
the shock ploughing through the outer iron core
might be small enough to allow a prompt explosion had to be
abandoned. The energy that is deposited behind the shock by the
outstreaming neutrinos
gained increased attention in light of a possible delayed shock revival
(Wilson \cite{Wilson_85}, Bethe and Wilson \cite{Bethe_Wilson_85}).
The amount of energy transfered, however, does not
alone depend on the neutrino spectrum and luminosity of the neutrinos,
but also on their angular distribution. The
neutrino propagation angle determines the path length in a given
radial shell; therefore, neutrinos with more tangential directions
have a higher chance of absorption. A known problem with the
multigroup flux-limited diffusion approximation to the Boltzmann
transport equation is that, in the important
semi-transparent region between the neutrinosphere and the
heating region, the angular distribution of the neutrinos is
not self-consistently determined by the transport
equation. The consequence is an underestimation of the
isotropy of the outstreaming neutrinos (Janka \cite{Janka_92}).

The focus therefore switched on the one hand to the
development of complete three-flavor Boltzmann neutrino transport
(Mezzacappa and Bruenn
\cite{Mezzacappa_Bruenn_93a,Mezzacappa_Bruenn_93b,Mezzacappa_Bruenn_93c}).
In stationary comparisons between full transport and multigroup
flux-limited diffusion approximations (Messer et al.
\cite{Messer_et_al_98}, Yamada et al. \cite{Yamada_Janka_Suzuki_99})
more efficient heating was found with accurate transport.
On the other hand, multidimensional phenomena were explored in
the hope of finding more efficient energy transfer by neutrinos.
The inclusion of neutron finger convection
(see also Ref. \cite{Bruenn_Dineva_96}) produced explosions
(Wilson and Mayle \cite{Wilson_Mayle_93}) and established the delayed
explosion scenario. Two-dimensional investigations of convection 
behind the shock and in the proto-neutron star followed
with mixed results (Herant et al.\cite{Herant_Benz_Colgate_92},
Miller et al.\cite{Miller_Wilson_Mayle_93},
Herant et al.\cite{Herant_et_al_94},
Burrows et al.\cite{Burrows_Hayes_Fryxell_95},
Janka and M\"uller \cite{Janka_Mueller_96},
Keil et al.\cite{Keil_Janka_Mueller_96},
Mezzacappa et al.\cite{Mezzacappa_et_al_98a,Mezzacappa_et_al_98b}).

Semi-analytical investigations (Burrows and Goshy \cite{Burrows_Goshy_93},
and recently Janka \cite{Janka_00}) illuminate
the basic mechanisms, but are not able to self-consistently decide
for or against explosions nor able to predict detailed data in either of
these cases.

In the search for a robust supernova mechanism, attention was directed
toward simulations in the nonrelativistic (NR) limit, because explosions
in GR seemed to be less likely in the selective picture that the
deeper shock formation
would produce larger dissociation losses and that the neutrino luminosities
would suffer gravitational redshift.
Multi-dimensional simulations that approximate general relativistic effects
were performed by Fryer \cite{Fryer_99} and Fryer and Heger
\cite{Fryer_Heger_00} who found reduced convection with
rotating progenitors.

However, there are also beneficial GR effects: Baron et al.
\cite{Baron_Cooperstein_Kahana_85b} reported prompt
explosions in general relativity with a very soft equation
of state and a leakage scheme, although these explosions
were not reproduced in a general relativistic multigroup
flux-limited diffusion simulation by Myra et al.
\cite{Myra_Bludman_89} when neutrino--electron scattering
was included. Swesty et al. \cite{Swesty_Lattimer_Myra_94}
also did not find an explosion in a systematic investigation of
realistic parameters in the equation of state. Late-time beneficial
effects from the decrease in the gravitational potential
by neutrino energy radiation were suggested by Goldman and
Nussinov \cite{Goldman_Nussinov_93}.
De Nisco et al. \cite{DeNisco_Bruenn_Mezzacappa_97}
in a quasi-static investigation pointed out
that, among many detrimental effects, the hotter core in general
relativistic
hydrodynamics produces higher neutrino luminosities with harder
spectra, resulting in a potentially beneficial impact on the heating
rate. Recent dynamical simulations with GR multigroup flux-limited
diffusion confirm this effect, although it does not appear to
be strong enough to outweigh the negative GR contributions,
i.e. the smaller heating region and greater infall velocities
(Bruenn et al. \cite{Bruenn_DeNisco_Mezzacappa_00}).

Recently, postbounce evolution was reexamined with Boltzmann
neutrino transport without invoking multidimensional effects.
As the result of
an undervalued nucleon isoenergetic scattering opacity, first
simulations led to an explosion of a $13$ M$_{\odot}$ progenitor
(Liebend\"orfer \cite{Liebendoerfer_00}). Simulations in Newtonian
gravity with $O(v/c)$ Boltzmann transport and standard nuclear
physics for the same progenitor do lead to an enhanced shock
radius, but not to an explosion
(Mezzacappa et al.\cite{Mezzacappa_et_al_00}). Even with the
omission of the energy loss due to the escape of $\mu$- and
$\tau$ neutrinos, independent simulations
of the postbounce evolution of a $15$ M$_{\odot}$ progenitor
do not produce an explosion (Rampp and Janka \cite{Rampp_Janka_00}).
The latter simulations are based on a tangent-ray method
for the $O(v/c)$ Boltzmann neutrino transport (see also
Burrows et al. \cite{Burrows_et_al_00}).

It is the purpose of this paper to report on the completion of a
general relativistic radiation hydrodynamics code for spherically
symmetric flows that solves the detailed Boltzmann transport equation,
and to present the self-consistent simulation
of stellar core collapse, bounce, and postbounce evolution
for a $13$ M$_{\odot}$ star, with all neutrino
flavors and ``standard'' nuclear physics included.


\section {\bf The Recipe}

Our simulation is initiated from the stellar precollapse model
of Nomoto and Hashimoto \cite{Nomoto_Hashimoto_88}. We use the
Lattimer-Swesty equation of state \cite{Lattimer_Swesty_91}.
Simplified silicon burning
is included as material from the silicon layer surrounding
the iron core is instantaneously burned under energy conservation
to nuclear statistical equilibrium as soon as the temperature
exceeds $0.44$ MeV (Mezzacappa et al. \cite{Mezzacappa_et_al_00}).
The simulations were carried out with a new general relativistic
neutrino radiation hydrodynamics code, AGILE-BOLTZTRAN, based on
a conservative formulation of GR radiation hydrodynamics
in spherical symmetry and comoving coordinates (Liebend\"orfer et al.
\cite{Liebendoerfer_Mezzacappa_Thielemann_00}).

AGILE is an implicit GR hydrodynamics code 
using an adaptive grid technique to conservatively implement
shift vectors (Liebend\"orfer and Thielemann 
\cite{Liebendoerfer_Thielemann_98}).
It maximizes, with respect to the number of required grid points,
the resolution in regions with large density gradients and
allows a smooth propagation of the shock through the outer layers
(see graph (a) in Figs. (\ref{1_bounce.ps}-\ref{5_fine.ps})).
BOLTZTRAN is an implicit three-flavor multigroup Boltzmann neutrino
transport solver (Mezzacappa and Bruenn
\cite{Mezzacappa_Bruenn_93a,Mezzacappa_Bruenn_93b,Mezzacappa_Bruenn_93c},
Mezzacappa and Messer \cite{Mezzacappa_Messer_98})
that was consistently coupled to AGILE, enabled for adaptive gridding,
and extended to GR flows under omission of the small gravitational
neutrino backreaction (Liebend\"orfer \cite{Liebendoerfer_00}).
We choose $103$ spatial zones, and discretize the neutrino-momentum phase
space, as in Mezzacappa et al. \cite{Mezzacappa_et_al_00}, with 
$6$-point Gaussian quadrature and $12$ neutrino energy groups,
ranging from $5$ MeV to $300$ MeV.

We emphasize at this point 
that the energy evolution in Lagrangian radiation hydrodynamics
obeys the very simple conservation equation
\begin{eqnarray}
\frac{\partial}{\partial t}\left[ \Gamma e +\frac{2}{\Gamma +1}\left( 
\frac{1}{2}u^{2}-\frac{m}{r}\right) +\Gamma J+uH\right]  &  & 
\nonumber \\
+\frac{\partial }{\partial a}\left[ 4\pi r^{2}\alpha
up+4\pi r^{2}\alpha \rho \left( uK+\Gamma 
H\right) \right]  & = & 
0.\label{eq_hydro+radiation_energy_conservation} 
\end{eqnarray}
The variables are defined as in  Lindquist \cite{Lindquist_66} with the
exceptions that \( J \), \( H \), and \( K \) represent the
zeroth, first, and second angular moments of the specific radiation energy;
\( \rho \), \( e \), and \( p \) the fluid rest mass density, internal energy,
and pressure; and, \( a \), \( r \), and \( u \) the enclosed rest mass, the
radius, and the time derivative of the radius devided by the lapse function
\( \alpha \).
The integration of Eq. 
(\ref{eq_hydro+radiation_energy_conservation}) over the total rest 
mass $a_{tot}=1.55$ M$_{\odot}$ in the computational domain provides a
check on the conservation of total energy, which can easily be monitored.
Energy conservation presents a significant challenge.
The cancellation of the gravitational binding energy with the
internal energy sets the scale of the
total energy, which is two orders of magnitude smaller than either
contribution separately and comparable to the
$10^{51}$ erg ($\sim1\%$ of radiation energy) expected to be
deposited in the heating region. 
Our total energy drift is no more than $5\times 10^{49}$ erg in
the first $200$ ms in which the explosion outcome is decided. Lepton
number is conserved to within a fraction of a percent.


\section{\bf The Failed Explosion of a $13$~M$_{\odot}$ Star}

\begin{figure}
\begin{center}
  \epsfig{file=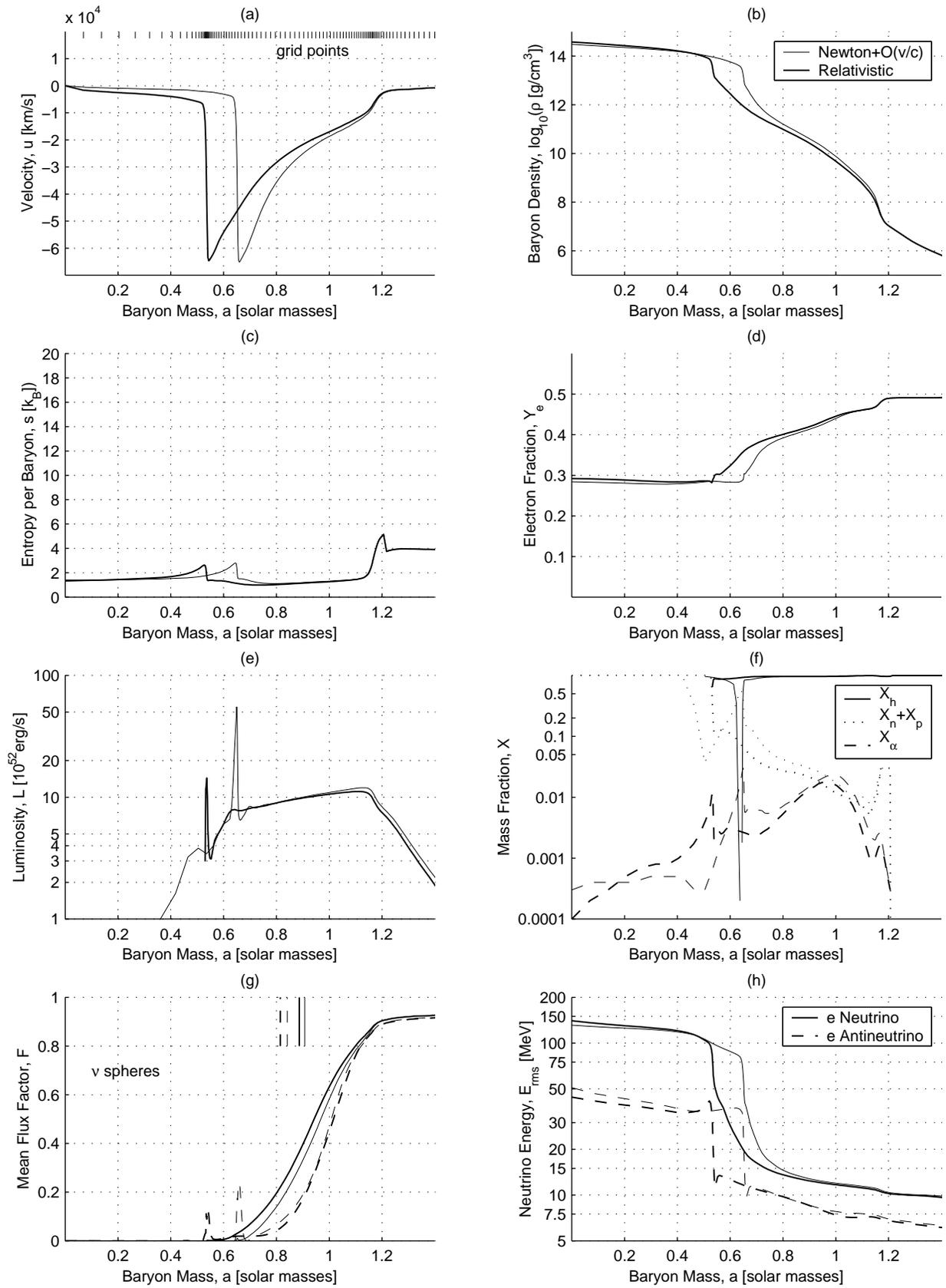,height=8.8in}
\end{center}
\caption{Shock formation: Time slice at bounce.}
\label{1_bounce.ps}
\end{figure}

\begin{figure}
\begin{center}
  \epsfig{file=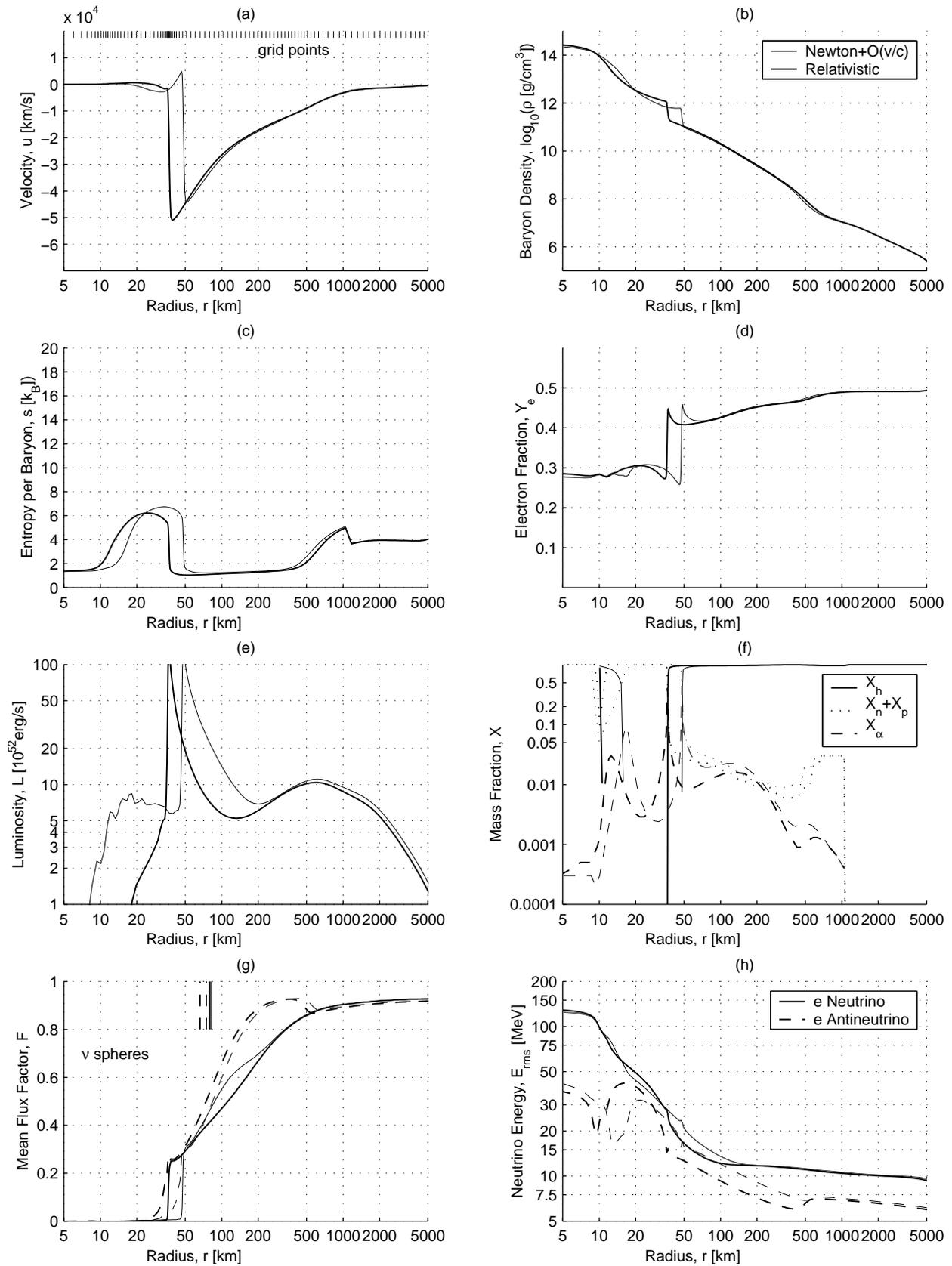,height=8.8in}
\end{center}
\caption{Dissociation: Time slice at $1$ ms after bounce.}
\label{2_dissociation.ps}
\end{figure}

\begin{figure}
\begin{center}
  \epsfig{file=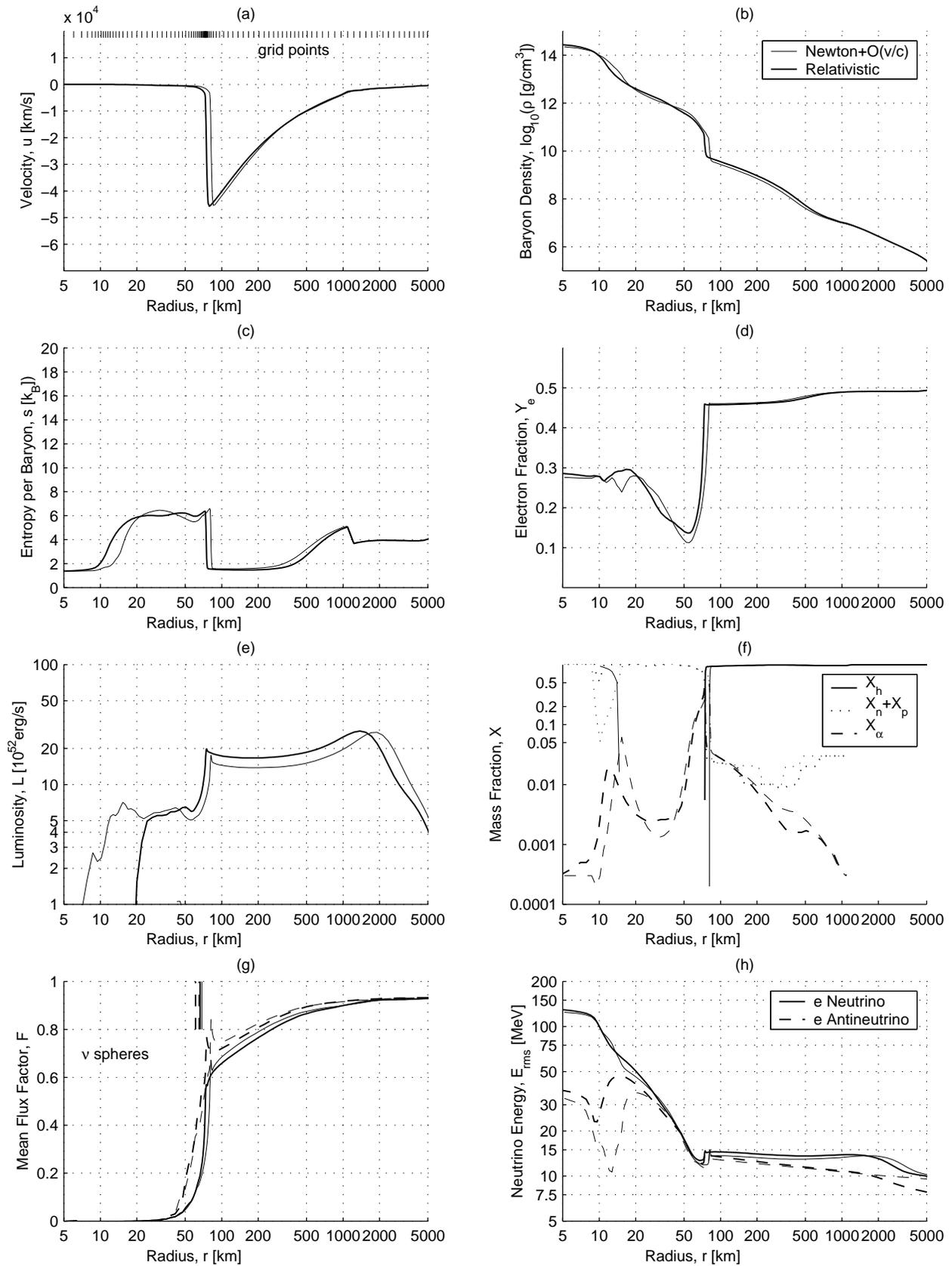,height=8.8in}
\end{center}
\caption{Neutrino burst: Time slice at $10$ ms after bounce.}
\label{3_burst.ps}
\end{figure}

\begin{figure}
\begin{center}
  \epsfig{file=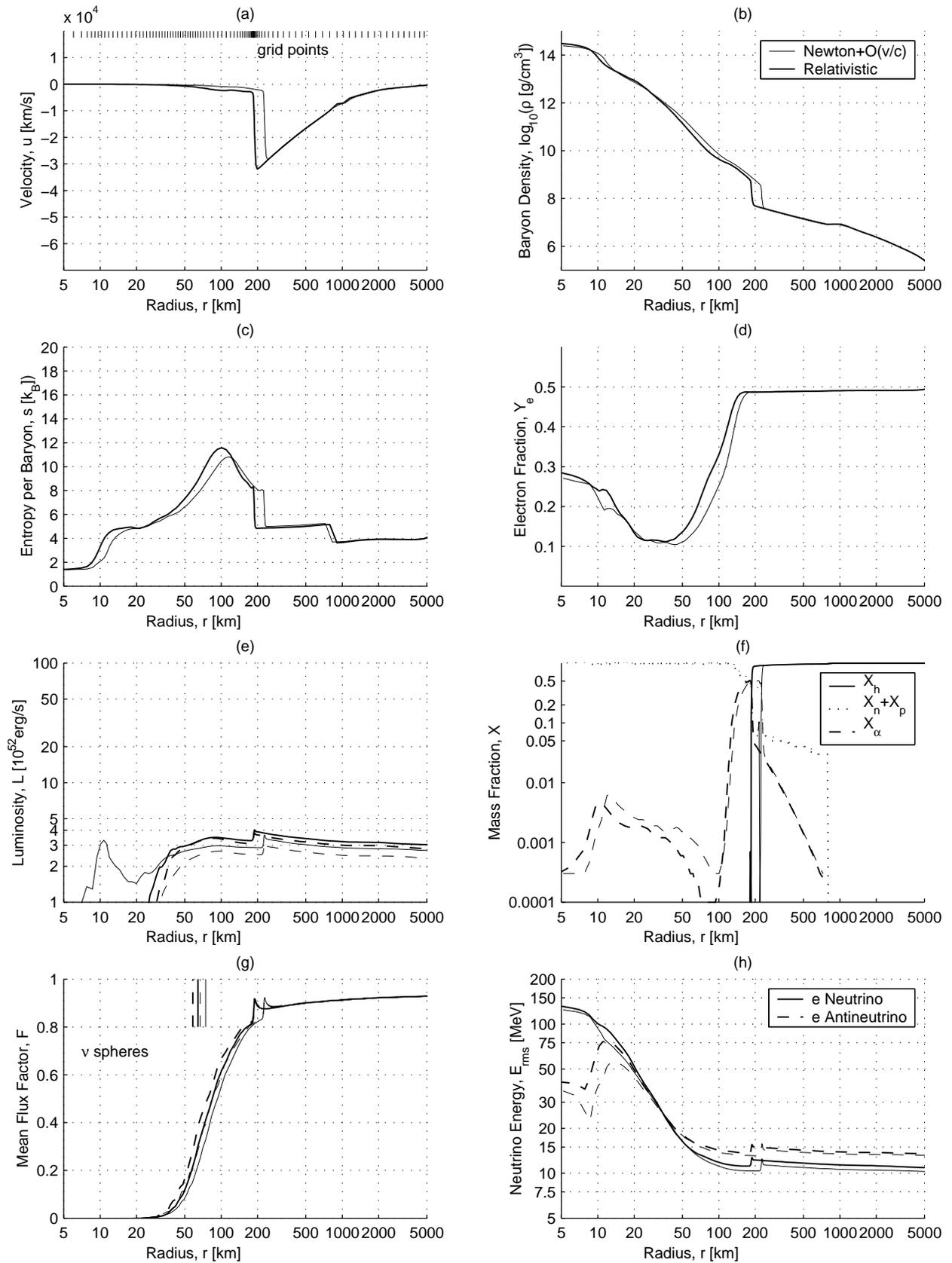,height=8.8in}
\end{center}
\caption{Neutrino heating: Time slice at $100$ ms after bounce.}
\label{4_heat.ps}
\end{figure}

\begin{figure}
\begin{center}
  \epsfig{file=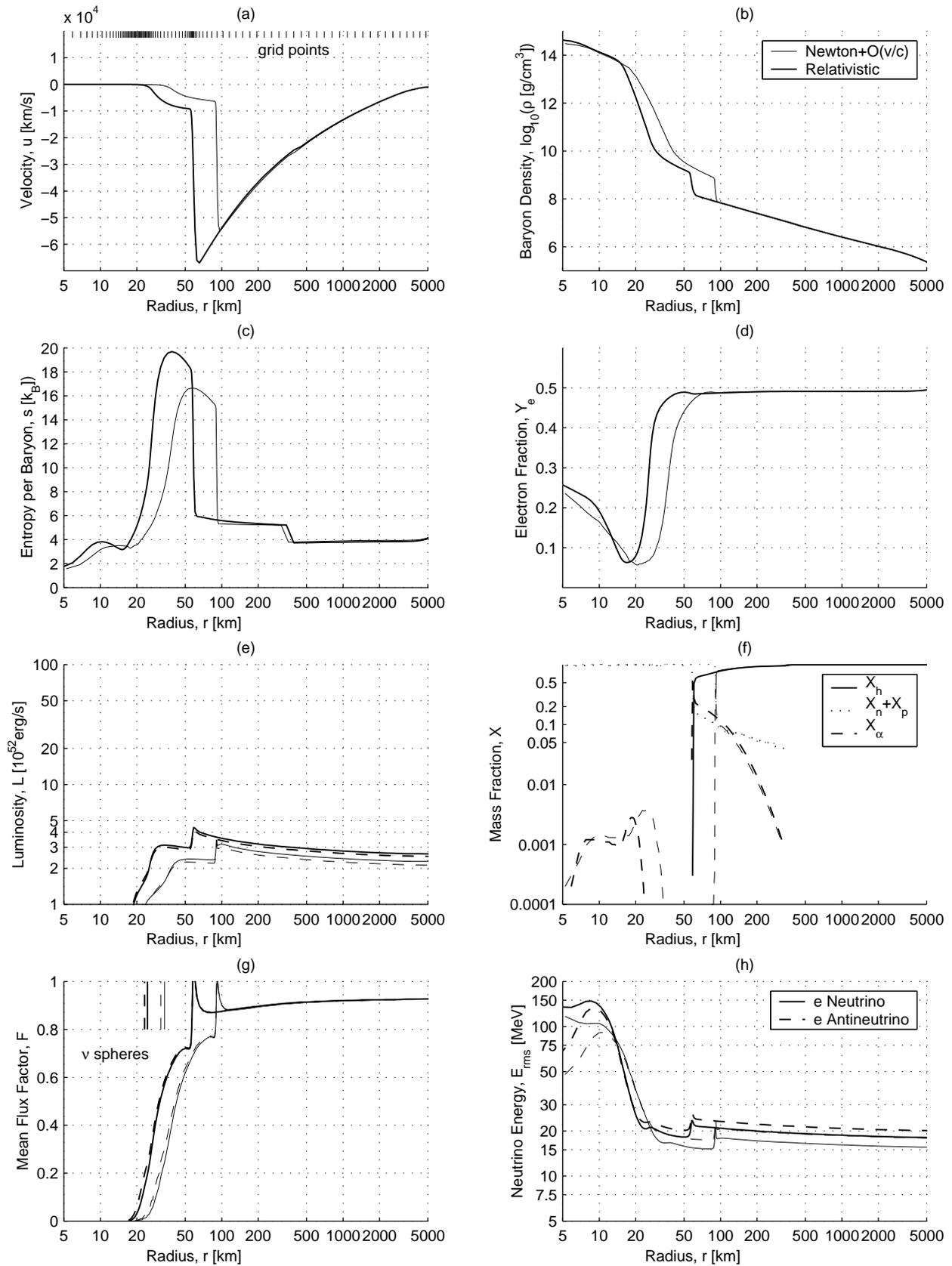,height=8.8in}
\end{center}
\caption{Shock recession: Time slice at $500$ ms after bounce.}
\label{5_fine.ps}
\end{figure}

We discuss the evolution of our simulation at five
time slices: At bounce, and at $1$ ms, $10$ ms, $100$ ms, and $500$ ms
after bounce. At
each time slice we show eight graphs that compare the general relativistic
evolution (GR, fat lines) to the nonrelativistic evolution
(NR, thin lines). We have already reported on the NR simulation
in Mezzacappa et al. \cite{Mezzacappa_et_al_00}.
The two runs are synchronized at bounce. The graphs
are (a) the velocity profile and the location of the adaptive
grid points (for the GR profiles only);
(b) the rest mass density profile; (c) the entropy
per baryon profile; (d) the electron fraction profile. The latter
three quantities enter the equation of state and determine the
composition and thermodynamic state of the local fluid element.
For the convenience of
interpretation, we additionally show the mass fraction of heavy
nuclei, free nucleons, and alpha particles in graph (f). The
remaining three graphs are devoted to neutrino radiation
quantities: (e) the luminosities and (h) the rms energies defined by
\begin{equation}
E_{rms} = \left( \frac{\int E^2 n_{\nu}(E) dE}{\int n_{\nu} (E) dE}
\right)^{\frac{1}{2}}.
\end{equation}
The integrand $n_{\nu}(E) dE$ is the number density of neutrinos
in the comoving frame energy interval $[E,E+dE]$.
Graph (g) shows
the mean flux factor defined by the quotient of the
neutrino energy flux, $H$, and the neutrino energy density,
$J$: $F\equiv H/(cJ)$.
We also show the energy-averaged neutrinospheres in this
plot because they separate the interior diffusion regime, with
small flux factors,
from the exterior regime, where large flux factors describe
increasingly forward-peaked neutrino propagation through
semitransparent and free-streaming regions.
AGILE-BOLTZTRAN solves directly for the specific neutrino
distribution function, $F(t,a,E,\mu)$, at each time $t$. The
angular resolution in the neutrino propagation angle cosine
$\mu$ is set by the order of a Gaussian quadrature. This limits the
representation of a strongly forward peaked radiation field, as
can be observed in the mean flux factor at large radii in the
graphs (g) of Fig. (\ref{1_bounce.ps}-\ref{5_fine.ps}). However,
our choice of 6-point quadrature allows an accurate representation of
the mean flux factor out to a radius of at least four times the
radius of the neutrinosphere. This range includes the cooling
and heating region for all times in the simulation, and any
influence of this numerical restriction on the dynamics is excluded.
To define an energy-averaged neutrinosphere,
we write the energy dependent optical depth at enclosed rest
mass, $a$, as
\begin{equation}
\tau (a,E) = \int_{a}^{a_{tot}}
\sum_{i}\frac{\sigma_{i}(E) }{4\pi r^2}
\frac{n_{i}}{\rho} da.
\label{eq_optical_depth}
\end{equation}
The variables are defined as in
Eq. (\ref{eq_hydro+radiation_energy_conservation}).
The index $i$ labels different reactions, with cross sections $\sigma_{i}$
and target number densities $n_{i}$. We draw in graph (g) the
energy-averaged neutrinospheres at the location \( r(a) \) where the
condition
\begin{equation}
\frac{\int\tau (a,E) n_{\nu} (a,E) dE}
{\int n_{\nu} (a,E) dE} = \frac{2}{3}
\end{equation}
for the enclosed mass $a$ holds.
The radiation quantities are given for the electron neutrino
and the electron antineutrino separately. The $\mu$ and $\tau$
neutrinos do not exchange energy with the heating region. Their
luminosities and rms energies are shown in later figures.
In our discussion, we follow the
general relativistic simulation and supply numbers for the
nonrelativistic case in parentheses.

The first prominent differences between the GR and NR simulations
appear at bounce in Fig. (\ref{1_bounce.ps}). At this
time, the inner core passes maximum compression with a
central density of $3.87\times 10^{14}$ g/cm$^{3}$
(NR: $3.05\times 10^{14}$ g/cm$^{3}$) as
shown in graph (b).
The homologous collapse of the causally connected
inner core is abruptly halted when the central region reaches nuclear
density. The short range nuclear forces suddenly increase
the stiffness of the equation of state. A pressure wave emerges
from the center and steepens into a shock wave at the border
of the inner iron core, where it faces the disconnectedly infalling
material from the outer iron core. We observe the shock formation
in the velocity profile in graph (a) at an enclosed mass of
$0.53$ M$_{\odot}$
(NR: $0.65$ M$_{\odot}$). The conversion of kinetic energy
into internal energy at the shock front becomes apparent in
the discontinuity of the entropy profile in graph (c).
The additional discontinuity at $\sim 1.2$ M$_{\odot}$ stems from
continued silicon burning by compressional heating in the
infalling material. From the composition in graph (f), we know
that the shock starts to dissociate the heavy nuclei as they fall
in. The shock in the NR case encloses a region of undissociated
heavy nuclei at high density. The extent of this region is
limited by the phase transition to bulk nuclear
matter in the Lattimer-Swesty equation of state. These heavy
nuclei are not present in the GR case
because the shock is formed at a smaller radius, almost coincident
with the location of the phase transition.
The electron fraction profile in graph (d)
exhibits values from $0.499$ in the outer layers down to
$0.29$ (NR: $0.28$) in the central region. The lepton
fraction is set by neutrino trapping during core collapse
(Sato \cite{Sato_75}, Mazurek \cite{Mazurek_76}).
Thereafter, smaller changes in the electron fraction
are allowed according to an establishing equilibrium
between electron- and neutrino-capture at constant lepton
fraction (see also Mezzacappa and Bruenn
\cite{Mezzacappa_Bruenn_93c}, Messer \cite{Messer_00}).
Determined by this equilibrium are the rms energies for the
electron neutrinos, as shown in graph (h). The mean
flux factor in graph (g) reflects an isotropic neutrino
distribution
in the innermost $0.7$ M$_{\odot}$ and outwardly directed
radiation at the border of the computational domain. The
peaks in the luminosity in graph (e) are caused by
numerical noise at the shock front. The small flux
factor in graph (g) for the electron neutrinos shows
that the short range energy flux at the peak is small
in the sense that it does not
exceed one percent of the prevailing neutrino energy
density multiplied by the velocity of light. An
influence on the dynamics is therefore excluded.

One millisecond later, in Fig. (\ref{2_dissociation.ps}), the
shock has moved from the edge of the homologous core to
densities $\sim 10^{12}$ g/cm$^{3}$ as shown in graph (b).
In graph (f) we observe a region of dissociated material behind
the shock. The heavy nuclei still present in the NR simulation
between the shock heated matter and bulk nuclear matter have
a mass of $0.15$ M$_{\odot}$. They leave the shock an additional
energy $\sim 2\times 10^{51}$ erg with respect to the GR
simulation. The NR shock
in graph (a) still contains kinetic energy, while the GR
shock has almost completely turned into an accretion
shock. In graph (g) we see that the shock approaches
the neutrinospheres. Plotted are the energy-averaged
neutrinospheres located in the
broad region where neutrinos with different energies
emanate from the diffusion regime. As the
shock passes this region at about $4$ ms after bounce,
an energetic neutrino burst will be released from the
hot shocked material, rendering it ``neutrino-visible'' to the
outside world. A rise in the electron neutrino
luminosity at the neutrinospheres at this time
is manifest in graph (e).
A dip in the rms energy profiles of the electron antineutrino
in graph (h) at a radius $\sim 11$ km results from the fact
that there are two distinct regions producing these neutrinos: 
the cold compressed center of the star and the
hot shocked mantle.

The time slice in Fig. (\ref{3_burst.ps}) corresponds to
$10$ ms after bounce, after the launch of the neutrino burst.
The luminosity peak in graph (e) has already propagated
to a radius of $1800$ km. This is consistent with the time of
shock breakout, $t=10$ ms - $1800$ km/$c$
$\simeq 4$ ms. The propagation of the neutrinos
is also visible in their breakout rms energies. In graph (h), the
neutrinos from collapse, with rms energies $\sim 10$ MeV,
are replaced with the burst neutrinos having an rms energy
of almost $15$ MeV. The
electron neutrinos in the burst were produced by
electron capture on free protons. The corresponding
deleptonization behind the shock is dramatic in
graph (d). The energy carried off with the neutrino burst
completely drains the shock in both the NR and the GR cases. The
pure accretion shock continues to propagate outwards
as infalling material is dissociated and layered on
the hot mantle (graph (a)). This stage is the definitive
end of a ``prompt,'' i.e., purely hydrodynamic, explosion.

In the standard picture, the ensuing evolution is driven
by electron neutrino heating. Electron flavor
neutrinos are diffusing out of the cold unshocked core
as well as created in the accreting and compactifying matter around the
neutrinospheres in a hot shocked mantle. On their escape from
the semi-transparent region, they deposit
energy behind the shock by absorption on free nucleons.
This situation is well established at $100$ ms after bounce,
corresponding to Fig. (\ref{4_heat.ps}).
At this time,
the accretion shock has passed the border of the original iron
core. In graph (a), it has stalled
at a radius of roughly $200$ km. The GR shock shows a smaller
stall radius than the NR shock, and the infall
velocity behind the shock is about two times larger than
in the NR case.
The incomplete dissociation of the infalling silicon
layer at the shock in graph (f) illustrates the weakening
of the shock. Half of the mass remains bound in alpha
particles when it crosses the shock and is dissociated
later during the fall through the heating
region. We additionally notice that the heavy nuclei
in the cold unshocked core have either been dissociated by
heating or have made the phase transition to bulk
nuclear matter. Thus, the same amount of mass has been
dissociated in the NR and GR cases and the corresponding
earlier advantage of the NR case in the energy balance
behind the shock does not persist.
We see in graph (g) that the neutrino radiation
becomes increasingly forward peaked outside of the
neutrinospheres. There is an extended
region between $50$ and $200$ km where the neutrinos
can deposit energy by absorption. In the interior
part of this region, the net cooling region, this process
is more than outweighed by neutrino emission of the
infalling material via electron and positron capture, as the material
is compressed, heated, and layered onto the proto-neutron star.
In the outer net heating region, however, electron flavor
neutrino absorption just exceeds neutrino emission,
leading to comparatively small net heating.
The heating can indirectly be observed in graph (c)
where the entropy per baryon starts to increase
behind the shock.
We read from graph (g) that the neutrinospheres
in the more compact and hotter GR core are at smaller
radii. This leads to emission of neutrinos
at higher rms energies (graph (h)). The discontinuity
in the rms energies at the shock are due to the
Doppler shift across it. All radiation quantities
are those measured by comoving
observers. This choice also leads to the luminosity
step in graph (e): the infalling observer heading
towards the center observes a higher luminosity than
the observer almost at rest behind the shock.
Although the GR luminosities in the heating region
are up to $20\%$ higher
than in the NR case, the heating is not more
effective in the GR case. The shocked material
settles onto the proto-neutron star with a higher
velocity in the deeper gravitational potential
and therewith
stays for a shorter time in the net heating region.

Finally, we show the situation at $500$ ms after
bounce in Fig. (\ref{5_fine.ps}). The shock in
graph (a) has receded to $60$ km in the GR case
and to $90$ km in the NR case. The accretion shock,
sitting deep in the gravitational potential,
experiences a high ram pressure from the infalling
material. All nuclei are completely dissociated
when they pass the shock (graph (f)). The infall
velocities behind the shock are as large
as $5000$ km/s in the NR simulation and almost
double that in the GR simulation. This produces
high accretion luminosities in the GR case,
as shown in graph (e). Higher
rms energies in the GR case are evident
in graph (h), as expected from the deeper position of the
neutrinospheres. The net
heating is not very efficient on the rapidly infalling
material, and the entropy per baryon increases
only very slowly at $\sim 1.7k_B$/($100$ ms). This
rise is due to the infall of higher entropy
material and the retraction of the shock
radius to deeper points in the gravitational
potential such that a larger kinetic energy from
the infalling material is dissipated. The entropy
per baryon reaches $20k_B$ in graph (c).
Overall, most features in the GR simulation
are enhanced relative to the NR simulation,
and the GR case exhibits a more compact structure in
the proto-neutron star and its surroundings.
The central density reached is $5.15\times 10^{14}$
g/cm$^3$ in graph (b).

\begin{figure}
\begin{center}
  \epsfig{file=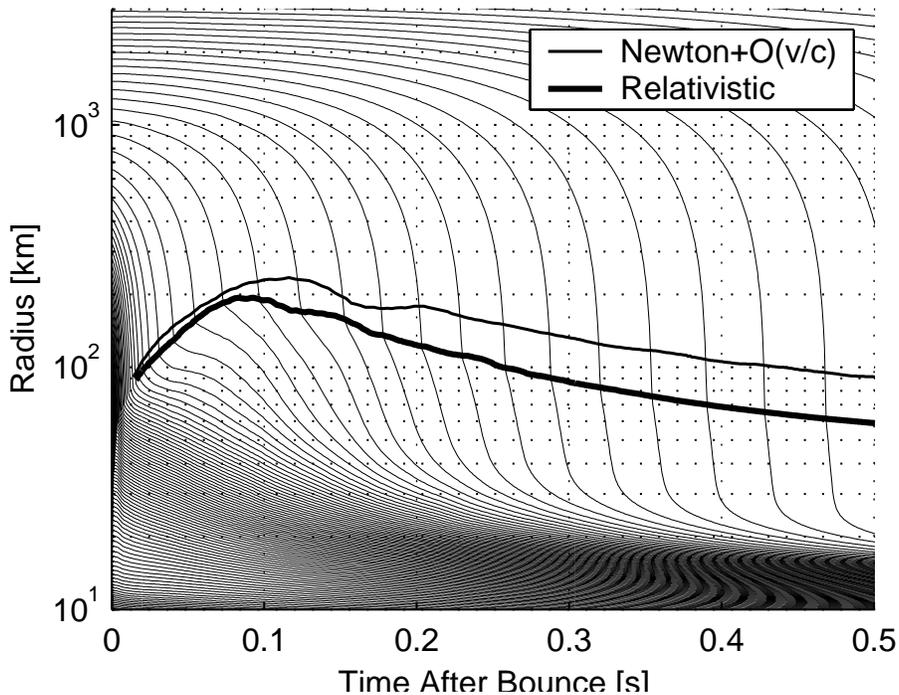,height=3.75in}
\end{center}
\caption{Radial trajectories of equal mass shells ($0.01$ M$_{\odot}$)
in the iron core and silicon layer.}
\label{6_traces.ps}
\end{figure}
The trajectories of mass shells containing $0.01$ M$_{\odot}$
are shown for the relativistic simulation in Fig. (\ref{6_traces.ps}).
Note that the high velocities
(steep gradients of mass traces) in the heating
region can hardly be distinguished from the ten times
larger infall velocities outside of the shock.
As discussed above, mass elements
are almost stopped at the shock front and do not
directly fall onto the proto-neutron
star surface, as might be inferred from Fig. (\ref{6_traces.ps}).
The shock position is drawn
for both the GR and the NR simulation. The GR shock barely
reaches $200$ km, the NR shock reaches $230$ km.

\begin{figure}
\begin{center}
  \epsfig{file=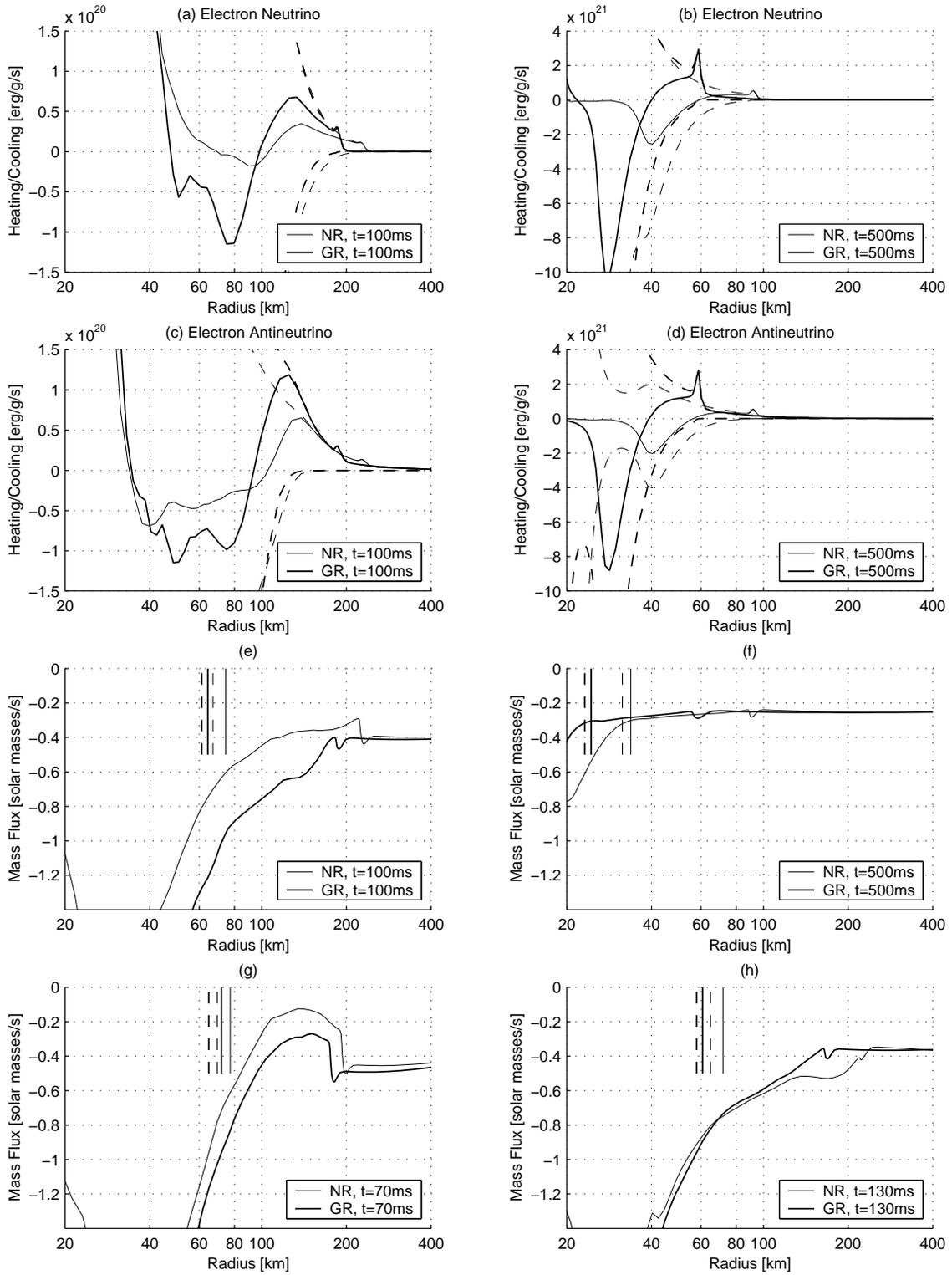,height=8in}
\end{center}
\caption{Graph (a) and (b) show the energy transfer rates from
emission and absorption of electron neutrinos at $100$ ms and $500$ ms
after bounce. The dashed lines mark the separate heating and cooling
rates whose superposition (solid line) leads to the net rate. Graphs
(c) and (d) show the same quantities for electron antineutrino emission
and absorption. Graphs (e)-(h) focus on the mass flux through spherical
shells observed at constant radius.
The vertical solid line shows the location
of the electron neutrinosphere, the dashed line marks the electron 
antineutrinosphere. The four graphs correspond to times $100$ ms
and $500$ ms, as above, and for the illustration of the time
dependence, to times $70$ ms and $130$ ms after bounce.}
\label{7_rates.ps}
\end{figure}

Graphs (a)-(d) of Fig. (\ref{7_rates.ps}) show the neutrino
heating rates (dashed lines with positive values), neutrino cooling
rates (dashed lines with negative values), and their superpositions
(solid lines). These energy transfer rates are shown separately
for the electron neutrinos (a,b)
and electron antineutrinos (c,d). We see that the net energy
transfer rate is the difference between much larger cooling
and heating rates, except very close to the shock, where heating
clearly dominates
the cooling. The location where the net heating changes sign
defines the gain radius that separates the net cooling
region (neutrinosphere to gain radius) from the net heating
region (gain radius to shock radius). The peak net heating rate
occurs rather close to the gain radius.
Note the sharp cutoff at the shock in the heating by electron
neutrino absorption with respect to the smoother decrease in electron
antineutrino absorption. This is due to the steep drop in the
free neutron mass fraction. Free neutrons are less abundant in
the presence of neutron rich heavy nuclei outside the
shock than the targets for electron antineutrino absorption,
the free protons.

At $100$ ms after bounce we observe a net heating rate
$\sim 10^{20}$ erg/g/s outside of the gain radius at
$\sim 100$ km. Note that the net heating by electron antineutrino
absorption is higher than the net heating by electron neutrino
absorption. This is due to the fact that the increasing electron
chemical potential for the infalling material, at comparable heating
rates, favors cooling by electron capture compared to cooling by positron
capture. This example illustrates that the net energy transfer or
the net change in the electron fraction by weak interactions
cannot be estimated simply based on luminosities and absorption cross
sections without a detailed consideration of the compensating inverse
reaction rates. We found increasing electron fractions that exceed a value
of $0.5$ only in our earlier exploding models where the
decrease in the chemical potential in the hot expanding
layers allowed for comparable electron and positron capture
at a slightly favored electron neutrino absorption rate
(e.g. in Ref. \cite{Liebendoerfer_00}).

Graph (e) shows the rest mass flux $4\pi r^2 u \rho/\Gamma$ through
spherical surfaces measured at constant radii $r$ at $100$ ms
after bounce. The time slices at $70$ ms, (g), and at
$130$ ms, (h), are added to sketch the time
dependence of the mass flux when the shock starts to recede.
The largest mass flux occurs at small radii where the high density
proto-neutron star slowly adjusts its size to the newly
accreted material. Most interesting for shock revival
is the region between the neutrinospheres (shown as vertical bars)
and the shock radius, which enclose the cooling and heating region
visible in graph (a) and (b).
The discontinuity in graph (g) in the mass flux
at the shock front indicates that the shock is still moving
outwards. The material at the gain radius and at the shock radius
show a comparable mass flux at $70$ ms after bounce. However,
at $100$ ms after bounce in graph (e), and even more so at $130$ ms
after bounce in graph (h), a slope in the mass flux develops. The
mass flux through the neutrinospheres is larger than the
mass flux through the gain radius, and both are
larger than the mass flux through the shock. The latter
is determined by the infalling material, i.e.,
the density profile of the progenitor and the gravitational
potential. The material in the heating region is drained from
below (Janka \cite{Janka_00}), and the conditions for heating
become inefficient because of the shortened time the infalling
matter spends in the heating region. In the general relativistic
simulation we observe this process earlier than in the
nonrelativistic simulation. The mass flux and the energy
transfer rates are about a factor of two larger.

At $500$ ms after bounce, the neutrinospheres and the shock have both
retracted to smaller radii. The mass flux in graph (f) is almost
constant
down to the neutrinospheres. It is mainly set by the close to free
infall of the layers around the iron core of the progenitor.
Although the infall velocities have reached very large values,
the actual mass flux has decreased because of the decreasing
density in the outer layers. The gain radius has adjusted to a much
smaller location $\sim 35$ km (NR $\sim 60$ km). In this compact
configuration, the energy transfer rates
have increased by more than an order of magnitude, as shown
in graphs (b) and (d). At this late-time, we observe
a peak in the heating rate across the shock front. Part of it
is due to the increased heating rate caused by the Doppler shift
in the rms energies and luminosities that applies to
the material in the shock front when it is
not yet completely decelerated but already dissociated.
Another part is caused by numerical artifacts in the
shock front: e.g., the unphysical width of the feature is
set by the artificial viscosity.
However, if we convolute the heating rate per mass with the
density decline in the shock, we find a smooth decrease in
the heating rate per volume across the shock and exclude relevant
perturbations in the results owing to this feature.

\begin{figure}
\begin{center}
  \epsfig{file=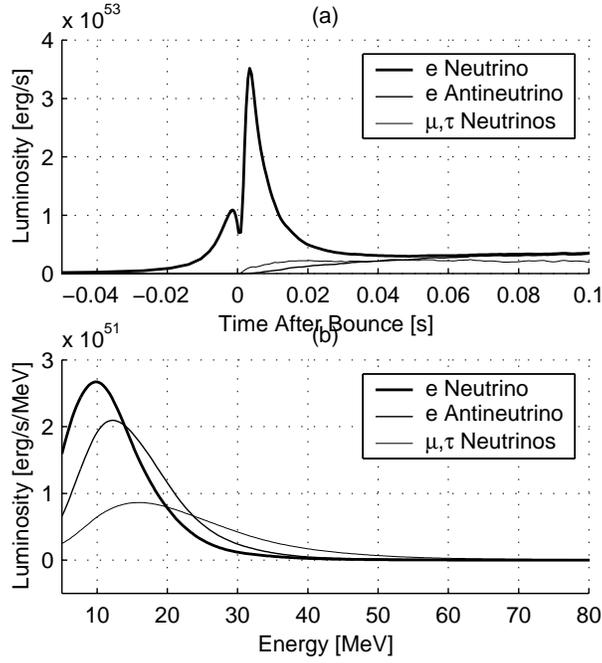,height=3.5in}
\end{center}
\caption{(a) Neutrino luminosities before and after bounce showing
the electron neutrino burst (sampled at $500$ km radius and time
shifted by $\Delta t=500$ km/$c$).
(b) Neutrino energy spectra at a radius
of $500$ km and $100$ ms after bounce.}
\label{8_spect.ps}
\end{figure}

\begin{figure}
\begin{center}
  \epsfig{file=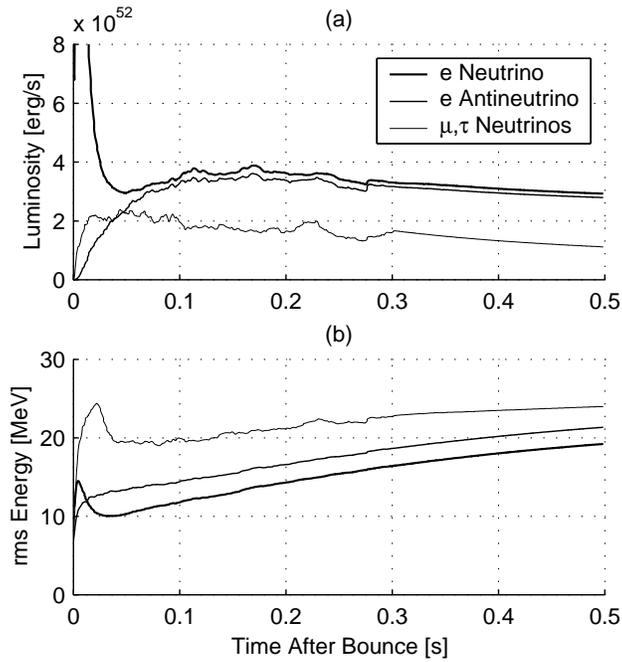,height=3.5in}
\end{center}
\caption{Neutrino luminosities and rms energies versus time at a 
radius of 500 km.}
\label{9_lumin.ps}
\end{figure}

We close the description of our simulations with two figures
showing the temporal evolution of the GR radiation
quantities. In Fig. (\ref{8_spect.ps}) we show
the neutrino luminosities at $500$ km radius as a
function of time.
The electron luminosity is slowly rising
during collapse and decreases as the core reaches
maximal density. It remains suppressed for the
$\sim 4$ ms the shock needs to propagate to the
electron neutrinosphere. The most prominent feature is
the electron neutrino burst, reaching $3.5\times 10^{53}$
erg/s at shock breakout, and declining afterwards.
Graph (b) shows the luminosity spectra at $100$ ms
after bounce, just before the shock starts to recede.
Figure (\ref{9_lumin.ps}) shows the long-term evolution
with better resolution, sampled at a radius of $500$ km.
The maximum luminosities in
the GR simulation are $3.9\times 10^{52}$ erg/s,
$3.6\times 10^{52}$ erg/s, and $2.3\times 10^{52}$ erg/s
for the electron neutrino, electron antineutrino,
and $\mu$ and $\tau$ neutrinos respectively. This is roughly
$10\%$ more than the maximum luminosities reached
(at different times) in the NR simulation.
The GR results also show
higher rms energies, but no qualitatively different
features.



\section{\bf Conclusion and Outlook}

We have completed the construction of a general
relativistic radiation hydrodynamics code, AGILE-BOLTZTRAN,
and simulated the spherically symmetric, general relativistic
core collapse, bounce, and postbounce evolution
of a $13$ M$_{\odot}$ progenitor. We investigate the
confluence of (i) matter and radiation in a deeper effective
gravitational potential,
(ii) a GR core hydrodynamic structure that acts as a
more intense neutrino source, and (iii) an increased
heating efficiency obtained from accurate three-flavor
Boltzmann neutrino transport. However, we find that the
combination of these ingredients does not result in a
supernova explosion. Our model shares
this outcome with recent simulations that investigated
a subset of these issues (Rampp and Janka
\cite{Rampp_Janka_00}, Mezzacappa et al. \cite{Mezzacappa_et_al_00},
Bruenn et al. \cite{Bruenn_DeNisco_Mezzacappa_00}).

(i) During core collapse, the homologous core is smaller in the
general relativistic gravitational potential than in Newtonian
gravity. This leads to a smaller enclosed mass at shock formation
and higher initial dissociation losses when the shock moves to larger
radii.
The neutrinos in curved spacetime propagate on trajectories
with nearly constant $\varepsilon=(\Gamma + u\mu)E$ and
$b=r\sqrt{1-\mu^2}/(\Gamma + u\mu)$, where $\mu$ and $E$ are the
neutrino propagation-angle cosine and energy measured by comoving 
observers. This latter effect is of little importance for our low-mass
progenitor. The GR corrections for redshift and curvature between
the neutrinospheres and the heating region do not exceed $3\%$
until shock recession, and increase only later, when the
neutrinospheres have receded to smaller radii.
(ii) Nonlinear effects in GR enhance the self-gravitation in the
high-density domain of the proto-neutron star. The
latter becomes more compact and exhibits higher internal energies.
This in turn leads to higher core and accretion luminosities 
with harder spectra in all neutrino flavors
(Bruenn et al.\cite{Bruenn_DeNisco_Mezzacappa_00}).
(iii) The solution of the Boltzmann equation for the neutrino 
transport reproduces accurately the angular distribution of
the neutrino radiation field behind the shock. An increased
angular spread in the semi-transparent regime
keeps the outstreaming neutrinos longer in the heating region and,
therefore, increases the absorption efficiency with respect to the
multigroup flux-limited diffusion approximation
(Messer et al.\cite{Messer_et_al_98}, Yamada et al.
\cite{Yamada_Janka_Suzuki_99}).

The main difference between the GR and NR simulations stems
from the difference in the size of the proto-neutron
star, which is caused by the nonlinear GR effects at very high
densities in the center of the
star. At radii of order $100$ km and larger,
the gravitational potential becomes comparable
in the GR and NR cases. However, large differences arise if the
steep gravitational well is probed at different
{\em positions} in the GR and NR simulations, as
happens with accretion down to the neutrinospheres,
which represent the outer ``boundary'' of the proto-neutron
star. The neutrinospheres in the more compact GR case receive
material that has traversed a larger potential difference
and, setteling deeper in the well, produces more energetic
accretion luminosities.
After the shock has stalled, the surrounding layers
(cooling-heating region, shock radius) adjust to a
smaller radius and settle to a stationary equilibrium
in the spirit of Burrows and Goshy \cite{Burrows_Goshy_93}.
In the GR simulation, this occurs at a smaller radius,
deeper in the gravitational well, with higher infall
velocities, higher accretion luminosities,
and higher heating rates, which sustain equilibrium
between the pressure behind the shock and the increased
ram pressure ahead of it.
The tight feedback between infall velocity,
accretion luminosity and heating leads to {\em higher} luminosities
in more compact configurations where the shock stalls at
a {\em smaller} radius. Moreover, this self regulation
has observable consequences: The gravitational well is
probed when the proto-neutron star radius is set, tuning
the directly observable neutrino luminosities and spectra.

From the comparison between the NR and GR shock trajectories
in Fig. (\ref{6_traces.ps}), we
might be tempted to conclude that the chances for a vigorous
explosion in the more realistic GR case are more
pessimistic than in the NR case. We however feel that such
a conclusion would be premature if only based on a postbounce
evolution that fails to reproduce an explosion.
An actual boost in the explosion energy due to general relativistic
effects has been seen by Liebend\"orfer \cite{Liebendoerfer_00} in
a simulation that led to an explosion because of incorrect
nucleon isoenergetic scattering opacities. As in the simulation
with the correct opacities, a more compact proto-neutron star
was created and higher luminosities with harder spectra
developed in the GR case. However, in the exploding case,
a turning point occurred where the infall velocities behind
the shock almost vanished. This happened $\sim 100$ ms after
bounce, before the heating region had the time to step deeper
into the gravitational well. The consequence was a more
efficient heating with the GR enhanced luminosities and spectra,
which lead to a more energetic supernova than in the corresponding
NR simulation. Our conclusion for low-mass progenitors with
small redshifts is:
GR effects do not cause explosions in our current simulations.
An essential physical ingredient is missing.
However, GR effects may eventually drive an explosion more
efficiently when it has been launched.

There are many facets of a supernova that cannot
consistently be included in a spherically symmetric model.
There is no doubt that convection behind the shock will
occur, and significant rotation and strong magnetic fields might
be present. Observations of neutron star kicks, mixing of species,
inhomogeneous ejecta, and polarization of spectra
suggest the presence of asymmetries in supernova explosions
(e.g. Tueller et al. \cite{Tueller_et_al_91},
Strom et al. \cite{Strom_et_al_95},
Galama et al. \cite{Galama_et_al_98},
Leonard et al. \cite{Leonard_et_al_00},
and references therein).
Motivated by such observations, various
multi-dimensional explosion mechanisms have been explored
(Herant et al.\cite{Herant_Benz_Colgate_92},
Miller et al.\cite{Miller_Wilson_Mayle_93},
Herant et al.\cite{Herant_et_al_94},
Burrows et al.\cite{Burrows_Hayes_Fryxell_95},
Janka and M\"uller \cite{Janka_Mueller_96},
Mezzacappa et al.\cite{Mezzacappa_et_al_98b},
Fryer \cite{Fryer_99},
Fryer and Heger \cite{Fryer_Heger_00}) and
jet-based explosion scenarios received new momentum
(H\"oflich et al. \cite{Hoeflich_Wheeler_Wang_99},
Khokhlov et al. \cite{Khokhlov_et_al_99},
MacFadyen and Woosley \cite{MacFadyen_Woosley_99},
Wheeler et al. \cite{Wheeler_et_al_00}).
It remains a challenge for the next decade
to include all relevant physics in supernova simulations
in an attempt to reproduce supernova observations in detail.
For the time being, one has to single out a subset of the
known physics and to investigate the role each part plays
in the restricted simulation. It is natural to start with
ingredients that have long been believed to be essential
for the explosion and to add modifiers in a systematic
way until the observables can be reproduced
(Mezzacappa et al. \cite{Mezzacappa_et_al_00}).
Spherically symmetric supernova
modeling has a long tradition and is nearing
a definitive point in the sense that
high-resolution hydrodynamics, general relativity,
complete Boltzmann neutrino transport, and reasonable nuclear
and weak interaction physics are being combined to dispel
remaining uncertainties. Models that are restricted to
spherical symmetry allow the detailed simulation and study of
radiation hydrodynamic flows in general relativity, with all
feedbacks included. Moreover, they will provide a gauge for
unavoidable approximations in future, more inclusive simulations.
The accurate description of the high-density region in the proto-neutron
star and realistic multigroup transport are prerequisits for the
evaluation of new developments in the input physics.
With the former well underway, research in spherically
symmetric supernova models can focus
on the uncertainties in the nuclear physics and neutrino interactions.
A major revision in the equation of state
(Shen et al. \cite{Shen_et_al_98}, see also Swesty et al.
\cite{Swesty_Lattimer_Myra_94}) or in the
neutrino-matter interactions (Burrows and Sawyer
\cite{Burrows_Sawyer_98}, Reddy et al. \cite{Reddy_et_al_99},
Thompson et al. \cite{Thompson_Burrows_Horvath_00},
Langanke and Martinez-Pinedo \cite{Langanke_Martinez_00}), or
the inclusion of neutrino oscillations
(Fuller et al. \cite{Fuller_et_al_92},
Abazajian et al. \cite{Abazajian_Fuller_Patel_01}),
might be the last opportunity to understand
the most basic supernova observable in restricted
spherical models: explosion.


\section*{Acknowledgments}
We enjoyed fruitful discussions with Thomas Janka and Markus Rampp.
M.L. is supported by the National Science Foundation under contract 
AST-9877130 and, formerly, was supported by the Swiss National 
Science Foundation under contract 20-53798.98. 
A.M. is supported at the Oak Ridge National Laboratory, managed by
UT-Battelle, LLC, for the U.S. Department of Energy under contract
DE-AC05-00OR22725. 
F.-K.T. is supported in part by the Swiss National Science Foundation 
under contract 20-61822.00 and as a Visiting Distinguished Scientist
at the Oak Ridge National Laboratory. 
O.E.B.M. is supported by funds from the Joint Institute for Heavy Ion 
Research and a Dept. of Energy PECASE award. 
W.R.H. is supported by NASA under contract NAG5-8405 and by funds 
from the Joint Institute for Heavy Ion Research and a Dept. of Energy
PECASE award.
S.W.B. is supported by the NSF under contract 96-18423 and by NASA under
contract NAG5-3903.
Our simulations were carried out on the ORNL Physics Division Cray J90 
and the National Energy Research Supercomputer Center Cray J90. 



\begin{thebibliography}{10}

\bibitem{Colgate_White_66}
S.~A. Colgate and R.~H. White, Astrophysical Journal {\bf 143},  626  (1966).

\bibitem{May_White_67}
M.~M. May and R.~H. White, Computational Physics {\bf 7}, 219 (1967).

\bibitem{Misner_Sharp_64}
C.~W. Misner and D.~H. Sharp, Physical Review {\bf 136}, B571 (1964).

\bibitem{Lindquist_66}
R.~W. Lindquist, Annals of Physics {\bf 37}, 487 (1966).

\bibitem{Wilson_71}
J.~R. Wilson, Astrophysical Journal {\bf 163},  209  (1971).

\bibitem{Baron_Cooperstein_Kahana_85a}
E. Baron, J. Cooperstein, and S. Kahana, Nuclear Physics A
{\bf 440}, 744 (1985).

\bibitem{Lattimer_Swesty_91}
J. Lattimer and F.~D. Swesty, Nuclear Physics {\bf A535},  331  (1991).

\bibitem{Tubbs_Schramm_75}
D. Tubbs and D. Schramm, Astrophysical Journal {\bf 201}, 467 (1975).

\bibitem{Schinder_Shapiro_82}
P.~J. Schinder and S.~L. Shapiro, Astrophysical Journal Supplement 
Series {\bf 50}, 23 (1982).

\bibitem{Bruenn_85}
S.~W. Bruenn, Astrophysical Journal Supplement Series {\bf 58}, 771 
(1985).

\bibitem{Arnett_77}
W.~D. Arnett, Astrophysical Journal {\bf 218},  815  (1977).

\bibitem{Bowers_Wilson_82}
R.~L. Bowers and J.~R. Wilson, Astrophysical Journal Supplement Series
{\bf 50}, 115 (1982).

\bibitem{Myra_et_al_87}
E.~S. Myra, S.~A. Bludman, Y. Hoffman, I. Lichtenstadt, N. Sack,
and K.~A. Van Riper, Astrophysical Journal {\bf 318}, 744 (1987).

\bibitem{Bruenn_89a}
S.~W. Bruenn, Astrophysical Journal {\bf 340}, 955 (1989).

\bibitem{Bruenn_89b}
S.~W. Bruenn, Astrophysical Journal {\bf 341}, 385 (1989).

\bibitem{Wilson_85}
J.~R. Wilson, in {\em Numerical Astrophysics}, edited by J.~M. Centrella,
J.~M. LeBlanc, and R.~L. Bowers (Jones and Bartlett, Boston, 1985).

\bibitem{Bethe_Wilson_85}
H.~A. Bethe and J.~R. Wilson, Astrophysical Journal {\bf 295}, 14 (1985).

\bibitem{Janka_92}
H.-T. Janka, Astronomy and Astrophysics {\bf 256}, 452 (1992).

\bibitem{Mezzacappa_Bruenn_93a}
A. Mezzacappa and S.~W. Bruenn, Astrophysical Journal {\bf 405},  669  (1993).

\bibitem{Mezzacappa_Bruenn_93b}
A. Mezzacappa and S.~W. Bruenn, Astrophysical Journal {\bf 405}, 637 (1993).

\bibitem{Mezzacappa_Bruenn_93c}
A. Mezzacappa and S.~W. Bruenn, Astrophysical Journal {\bf 410},  740  (1993).

\bibitem{Messer_et_al_98}
O.~E.~B. Messer, A. Mezzacappa, S.~W. Bruenn, and M.~W. Guidry, Astrophysical
  Journal {\bf 507},  353  (1998).

\bibitem{Yamada_Janka_Suzuki_99}
S. Yamada, H.-T. Janka, and H. Suzuki, Astronomy and Astrophysics {\bf 344},
  533  (1999).

\bibitem{Bruenn_Dineva_96}
S.~W. Bruenn and T. Dineva, Astrophysical Journal Letters {\bf 458}, 
L71 (1996).

\bibitem{Wilson_Mayle_93}
J.~R. Wilson and R.~W. Mayle, Physics Reports {\bf 227},  97  (1993).

\bibitem{Herant_Benz_Colgate_92}
M. Herant, W. Benz, and S.~A. Colgate, Astrophysical Journal {\bf 395},  642
  (1992).

\bibitem{Miller_Wilson_Mayle_93}
D.~S. Miller, J.~R. Wilson, and R.~W. Mayle, Astrophysical Journal {\bf 415},
  278  (1993).

\bibitem{Herant_et_al_94}
M. Herant, W. Benz, R.~W. Hix, C.~L. Fryer, and S.~A. Colgate,
Astrophysical Journal {\bf 435},  339  (1994).

\bibitem{Burrows_Hayes_Fryxell_95}
A. Burrows, J. Hayes, and B.~A. Fryxell, Astrophysical Journal {\bf 450},  830
  (1995).

\bibitem{Janka_Mueller_96}
H.-T. Janka and E. M\"{u}ller, Astronomy and Astrophysics {\bf 306},  167
  (1996).

\bibitem{Keil_Janka_Mueller_96}
W. Keil, H.-T. Janka, and E. M\"uller, Astrophysical Journal Letters
{\bf 473}, L111 (1996).

\bibitem{Mezzacappa_et_al_98a}
A. Mezzacappa, A.~C. Calder, S.~W. Bruenn, J.~M. Blondin, M.~W. 
Guidry, M.~R. Strayer, and A.~S. Umar, Astrophysical Journal
{\bf 493},  848 (1998).

\bibitem{Mezzacappa_et_al_98b}
A. Mezzacappa, A.~C. Calder, S.~W. Bruenn, J.~M. Blondin, M.~W. 
Guidry, M.~R. Strayer, and A.~S. Umar, Astrophysical Journal
{\bf 495},  911  (1998).

\bibitem{Burrows_Goshy_93}
A. Burrows and J. Goshy, Astrophysical Journal Letters {\bf 416},  L75  (1993).

\bibitem{Janka_00}
H.-T. Janka, astro-ph/0008432, Astronomy and Astrophysics (to be published),
(2001).

\bibitem{Fryer_99}
C.~F. Fryer, Astrophysical Journal {\bf 522}, 413 (1999).

\bibitem{Fryer_Heger_00}
C.~F. Fryer and A. Heger, Astrophysical Journal {\bf 541}, 1033 (2000).

\bibitem{Baron_Cooperstein_Kahana_85b}
E. Baron, J. Cooperstein, and S. Kahana, Physical Review Letters {\bf 55},
126 (1985).

\bibitem{Myra_Bludman_89}
E.~S. Myra and S.~A. Bludman, Astrophysical Journal {\bf 340}, 384 (1989).

\bibitem{Swesty_Lattimer_Myra_94}
F.~D. Swesty, J.~M. Lattimer, and E.~S. Myra, Astrophysical Journal
{\bf 425}, 195 (1994).

\bibitem{Goldman_Nussinov_93}
I. Goldman and S. Nussinov, Astrophysical Journal {\bf 403}, 706 (1993).

\bibitem{DeNisco_Bruenn_Mezzacappa_97}
K. De Nisco, S.~W. Bruenn, and A. Mezzacappa, 
Bulletin of the 191st American Astronomical Society Meeting, Washington
DC, 1998, edited by R.~W. Milkey {\bf 29}, 191, 39.10 (AAS, Washington, 1997).

\bibitem{Bruenn_DeNisco_Mezzacappa_00}
S.~W. Bruenn, K.~R. DeNisco, and A. Mezzacappa, astro-ph/0101400,
Astrophysical Journal (to be published), (2001).

\bibitem{Liebendoerfer_00}
M. Liebend\"{o}rfer, Ph.D. thesis, University of Basel, (Basel, Switzerland,
2000).

\bibitem{Mezzacappa_et_al_00}
A. Mezzacappa, M. Liebend\"orfer, O.~E.~B. Messer, R.~W. Hix, F.-K. 
Thielemann, and S.~W. Bruenn, Physical Review Letters {\bf 86}, 1935 (2001).

\bibitem{Rampp_Janka_00}
M. Rampp and H.-T. Janka, Astrophysical Journal Letters {\bf 539},
L33 (2000).

\bibitem{Burrows_et_al_00}
A. Burrows, T. Young, Ph. Pinto, R. Eastman, and T.~A. Thompson,
Astrophysical Journal {\bf 539}, 865 (2000).

\bibitem{Nomoto_Hashimoto_88}
K. Nomoto and M. Hashimoto, Physics Reports {\bf 163},  13  (1988).

\bibitem{Liebendoerfer_Mezzacappa_Thielemann_00}
M. Liebend\"orfer, A. Mezzacappa, and F.-K. Thielemann,
Physical Review D {\bf 63}, 104003 (2001).

\bibitem{Liebendoerfer_Thielemann_98}
M. Liebend\"{o}rfer and F.-K. Thielemann,  in {\em Nineteenth Texas Symposium
  on Relativistic Astrophysics}, edited by E. Aubourg, T. Montmerle, J. Paul,
  and P. Peter (Elsevier Science B.~V., Amsterdam, 2000).

\bibitem{Mezzacappa_Messer_98}
A. Mezzacappa and O.~E.~B. Messer, Journal of Computational and Applied
  Mathematics {\bf 109},  281  (1999).

\bibitem{Sato_75}
K. Sato, Progr. Theoret. Phys. {\bf 53}, 595, and {\bf 54}, 1325 (1975).

\bibitem{Mazurek_76}
T. Mazurek, Astrophysical Journal {\bf 207}, 187 (1976).

\bibitem{Messer_00}
O.~E.~B. Messer, Ph.D. thesis, University of Tennessee, (Knoxville, USA, 2000).

\bibitem{Tueller_et_al_91}
J. Tueller, S. Barthelmy, N. Gehrels, M. Leventhal, C.~J. MacCallum,
and B.~J. Teegarden, in {\em Supernovae}, edited by S.~E. Woosley,
278 (Springer, Berlin, 1991).

\bibitem{Strom_et_al_95}
R. Strom., H.~M. Johnston, F. Verbunt, and B. Aschenbach,
Nature (London) {\bf 373}, 587 (1995).

\bibitem{Galama_et_al_98}
T.~J. Galama, P.~M. Vreeswijk, J. van Paradijs, C. Kouveliotou,
T. Augusteijn, O.~R. Hainaut, F. Patat, H. Boehnhardt, J. Brewer,
V. Doublier, J.-F. Gonzalez, C. Lidman, B. Leibundgut, J. Heise,
J. in 't Zand, P.~J. Groot, R.~G. Strom, P. Mazzali, K. Iwamoto,
K. Nomoto, H. Umeda, T. Nakamura, T. Koshut, M. Kippen, C. Robinson,
P. de Wildt, R.~A.~M.~J. Wijers, N. Tanvir, J. Greiner, E. Pian,
E. Palazzi, F. Frontera, N. Masetti, L. Nicastro, E. Malozzi, M. Feroci,
E. Costa, L. Piro, B.A. Peterson, C. Tinney, B. Boyle, R. Cannon,
R. Stathakis, M.C. Begam, P. Ianna,
Nature {\bf 395}, 670 (1998).

\bibitem{Leonard_et_al_00}
D.~C. Leonard, A.~V. Filippenko, A.~J. Barth, and T. Matheson,
Astrophysical Journal {\bf 536}, 239 (2000).

\bibitem{Hoeflich_Wheeler_Wang_99}
P. H\"oflich, J.~C. Wheeler, and L. Wang,
Astrophysical Journal {\bf 521}, 179 (1999).

\bibitem{Khokhlov_et_al_99}
A.~M. Khokhlov, P.~A. H\"oflich, E.~S. Oran, J.~C. Wheeler, L. Wang, and
A.~Yu. Chtchelkanova, Astrophysical Journal Letters {\bf 524}, 107 (1999).

\bibitem{MacFadyen_Woosley_99}
A.~I. MacFadyen and S.~E. Woosley, Astrophysical Journal {\bf 524},
262 (1999).

\bibitem{Wheeler_et_al_00}
J.~C. Wheeler, I. Yi, P. H\"oflich, and L. Wang, Astrophysical Journal {\bf 537},
810 (2000).

\bibitem{Shen_et_al_98}
H. Shen, H. Toki, K. Oyamatsu, and K. Sumiyoshi,
Nuclear Physics A {\bf 637}, 435 (1998).

\bibitem{Burrows_Sawyer_98}
A. Burrows and R.~F. Sawyer, Physical Review C {\bf 58}, 554 (1998).

\bibitem{Reddy_et_al_99}
S. Reddy, M. Prakash, J.~M. Lattimer, and J.~A. Pons, Physical Review
C {\bf 59}, 2888 (1999).

\bibitem{Thompson_Burrows_Horvath_00}
T.~A. Thompson, A. Burrows, and J.~E. Horvath, Physical Review C {\bf 62},
035802 (2000).

\bibitem{Langanke_Martinez_00}
K.~H. Langanke and G. Martinez-Pinedo, Nuclear Physics A {\bf 673},
481 (2000).

\bibitem{Fuller_et_al_92}
G.~M. Fuller, R.~W. Mayle, B.~S. Meyer, and J.~R. Wilson,
Astrophysical Journal {\bf 389}, 517 (1992).

\bibitem{Abazajian_Fuller_Patel_01}
K. Abazajian, G.~M. Fuller, and M. Patel, astro-ph/0101524 (2001).

\end{thebibliography}
\end{document}